\documentstyle[12pt,epsfig,aps]{revtex}

\begin{document}
\vspace{0.3truecm}
\begin{center}
{\Large
Inverse problem and Darboux transformations for two-dimensional
finite-difference Schr\"odinger equation }
\vspace{0.5cm}

      		{\large A.A. Suzko} \footnote{
Radiation Physics
and Chemistry Problems Institute, Academy of Sciences of Belarus, Minsk}
\end{center}
\vspace{0.3cm}
~~~~~~~~~~~~~{\large  Joint Institute for Nuclear Research, Dubna}
\vspace{0.5cm}

\begin{abstract}
A discrete version of the two-dimensional inverse scattering problem
is considered. On this basis, algebraic transformations for the
two-dimensional finite-difference Schr\"odinger equation are elaborated.
Generalization of the technique of one-dimensional Darboux
transformations for the two-dimensional finite-difference Schr\"odinger
equation is presented. Ideas of Bargmann-Darboux transformations for
the differential one-dimensional multichannel Schr\"odinger equation
are used for the two-dimensional lattice Schr\"odinger equation.
Analytic relationships are established between
different discrete potentials and the corresponding solutions.
\end{abstract}

\section{Introduction}
Study of multidimensional and multi-particle objects is qualitatively
more complicated than that of the one-dimensional case. When there is
no symmetry in the potentials of interaction between particles, these systems are
described by partial differential equations with unseparable variables.
Just for this reason the inverse scattering problem in 2- and 3-dimensional
spaces was formulated by Faddeev \cite{fadd} and Newton \cite{newt,newt2},
Novikov and Henkin \cite{novikov} at a much later than the 1-dimensional
problem. The foundation for developing the 2-dimensional inverse problem
in finite differences was laid by Berezanskii \cite{berez} who developed
the theory of orthogonal polynomials for the Jacobi infinite matrix. The
Darboux transformations in quantum mechanics for the Schr\"odinger
differential and difference 1-dimensional equation have much in common
with the spectral transformations for orthogonal polynomials \cite{spirid}.
It is therefore expedient to analyze the spectral inverse problem for the
discrete 2-dimensional Schr\"odinger equation on the basis of the
technique of orthogonalization of polynomials. The formulae of
the inverse problem with degenerate kernels are closely related to
Bargmann-Darboux transformations \cite{book}--\cite{suz2}.
The one-dimensional Darboux transformations has already found
wide applications in quantum mechanics and in the theory of nonlinear
integrable systems \cite{matv,matv1,spirid,spirid1}.

In the present paper, a discrete version of the Gelfand-Levitan inverse
spectral problem is considered for the two-dimensional lattice
Schr\"odinger equation. The two-dimensional finite-difference
inverse problem, based on the procedure of orthogonalization of polynomial
vectors, is a generalization of the one-dimensional procedure given in
\cite{case}. On the basis of the obtained formulae of the inverse problem,
the relations are derived for discrete Bargmann-Darboux transformations
in two dimensions.
The suggested algebraic approach allows one to construct families of
discrete potentials in an explicit form and the corresponding solutions.

\section{Inverse problem }

Consider the Schr\"odinger equation whose Hamiltonian is
tridiagonal in a certain basis
with respect to both the coordinate variables $n$ and $m$
\begin{eqnarray}
\label{1.1}
&& (H\psi)_{nm} = a_{n m}\psi(n-1,m)+a_{n+1 m}\psi(n+1,m) \nonumber\\
&&+b_{n m}\psi(n,m-1)+b_{n m+1}\psi(n,m+1)+c_{nm}\psi(n,m)=\lambda\psi(n,m).
\end{eqnarray}
The coefficients $a_{n m}, b_{n m}, c_{n m}$ are assumed to
be real and represent discrete potentials; $\psi(n,m)$ are discrete wave
functions, $(n, m)$ is an integer point of the half-plane,
$n=0,1,2,.., m=...,-1,0,1,...$, $\lambda$ is a spectral parameter.
The index $n$ can vary from $0$ to $\infty$ and the index $m$
can vary from $-\infty$ to $\infty$ or from $0$ to $\infty$; when
$0\leq n\leq  N$ and $0\leq m\leq M$, it is a special class of restricted
problems. The equation in finite differences (\ref{1.1}) can be represented
by the expression with operator coefficients
\begin{eqnarray}
\label{1.2}
(H\Psi)_n = A_{n}\Psi(n-1)+V_{n}\Psi(n)+A_{n+1}\Psi(n+1) =
\lambda \Psi(n),
\end{eqnarray}
when we treat one of the variables, $n$,
as the only discrete coordinate; and the other, $m$, as the channel index.
If we set $\Psi(-1)=0$, the action of the Schr\"odinger discrete operator
$H$ on the vector $\Psi=\{\Psi(0),\Psi(1),\Psi(2),...\Psi(n),...\}$
is represented by the action of the Jacobi block matrix $J$ on $\Psi$
\begin{eqnarray}
\label{1.J}
(J\Psi)_n=\left(\begin{array}{ccccccc}
V_{0} & A_{1}    &0     & 0    & 0      &...       & 0\\
A_{1} & V_{1}    & A_{2}& 0    & 0      &...       & 0 \\
0     & A_{2}    & V_{2}& A_{3}& 0      &...       & 0  \\
0     &  0       &  0   &  .   &  .     & .        & 0 \\
    . &  .       &      .& 0   & A_{n} & V_{n}   &A_{n+1}\\
    .&.          &.      & .        &     .    &. & .
\end{array}\right)
\left(\begin{array}{c}\Psi(0)\\ \Psi(1)\\ \Psi(2)\\.\\
\Psi(n)\\.
 \end{array}\right)=
\lambda
\left(\begin{array}{c}\Psi(0)\\ \Psi(1)\\ \Psi(2)\\.\\
\Psi(n)\\.
 \end{array}\right),
\end{eqnarray}
whose elements $V_n$ and $A_n$ are matrices at each fixed $n$,
and each element $\Psi(n)$ of the vector $\Psi$
corresponds to the vector $\{\psi_m(n)\},\psi_m(n)\equiv
\psi(n,m)$, in the other space variable $"m"$
\begin{eqnarray}
\label{1.V}
V_n=\left(\begin{array}{cccccccc}
   c_{n0}&b_{n1} &0      &0         &    0     &...       & 0 &.\\
b_{n1}& c_{n1}   &b_{n2} &0         &0         &...       & 0 &.\\
    0 & b_{n2}   & c_{n2}& b_{n3}   & 0        &...       & 0 &.\\
.     & .        & .     &  .       & .        &...       & 0 &.\\
    . &.         &      .& 0        & b_{nm-1} & c_{nm-1}&b_{nm}&.\\
    .&.          &.      & .        &     0    &b_{nm} & c_{nm}&b_{nm+1} \\
    .&.          &.      & .        &     .  &. & .&.
\end{array}\right),
\end{eqnarray}
$$A_n=\left(\begin{array}{cccccc}
a_{n0}& 0     & 0 & ...& 0\\
0     & a_{n1}& 0 & ...& 0\\
0     & 0     & a_{n2}& ...& 0\\
.     & .     & .     & .  & . \\
0     &  0    & 0     & ...& a_{nm}\\
.     & .     & .     & .  & .
\end{array}\right),~~
\Psi(n)=\left(\begin{array}{c}\psi_{0}(n)\\ \psi_{1}(n)\\  \psi_{2}(n)\\.\\
\psi_{m}(n)\\.
 \end{array}\right).
$$
It is clear that the operator $H$ is Hermitian. Note that the coupling
of equations (\ref{1.1}) or (\ref{1.2}) with respect to $m$ occurs only
between immediate neighbours and for this reason the symmetric coupling
matrix $V_{n;(mm')}\equiv V_{mm'}(n)$ in (\ref{1.2}) is tridiagonal and
relates the functions $\psi(n,m)$ at $m, m\pm 1$, unlike the conventional
multichannel case.

From the Jacobi block matrix (\ref{1.J}) tridiagonal in the variable $n$
it is seen that every vector of the solutions $\Psi(n)$ is connected with
the vectors of solutions $\Psi(n\pm 1)$ at the neighboring points
$ n\pm 1 $. As a result, if we take nonhomogeneous boundary conditions at
one end of the interval $ 0 \leq n \leq N $, we can obtain the solutions
on the whole interval by moving by subsequent steps from that end.
It turns out that these vectors of solutions as
functions of the spectral parameter $\lambda $ are polynomials
of $\lambda $ with matrix coefficients \cite{berez}  which
can be orthogonalized in the spectral measure.

{ The spectral inverse
problem is reduced to the construction of potential matrices $V_n$, $A_n$
and an unknown system of orthonormal polynomials with the use of the
known system of orthonormal polynomials corresponding to the
finite-difference equation (\ref{1.2}) but with the known
matrices $\stackrel{\circ}{V}_n$ and $\stackrel{\circ}{A}_n$.}

\subsection{Orthogonalization of polynomials}

Introduce auxiliary solutions $\varphi_{ms}(n)\equiv\varphi_s(\lambda,n,m) $
and $\stackrel{\circ}{\varphi}_{ms}(n)$ to eq.(\ref{1.1}) with the help
of the boundary conditions
\begin{eqnarray}
\label{1.3}
\varphi_{ms}(-1) =  0;~~
\varphi_{ms}(0)=\delta_{ms}; \\
\stackrel{\circ}{\varphi}_{ms}(-1) =  0;~~
\stackrel{\circ}{\varphi}_{ms}(0)=\delta_{ms},\nonumber
\end{eqnarray}
where $s$ is a point on the $m$ axis. The solutions
$\stackrel{\circ}{\varphi}_{s} $ satisfy the same equation (\ref{1.1})
with the known potentials $\stackrel{\circ}{a}_{nm},
\stackrel{\circ}{c}_{nm}=\stackrel{\circ}{V}_{mm}(n),
\stackrel{\circ}{b}_{nm+1}=\stackrel{\circ}{V}_{mm+1}(n), $
$\stackrel{\circ}{b}_{nm}=\stackrel{\circ}{V}_{mm-1}(n)$
\begin{eqnarray}
\label{1.1a}
\stackrel{\circ}{a}_{nm}\stackrel{\circ}{\varphi}_{ms}(\lambda,n-1)+
\stackrel{\circ}{a}_{n+1 m}\stackrel{\circ}{\varphi}_{ms}(\lambda,n+1)
+\sum_{m''=m-1}^{m+1}\stackrel{\circ}{V}_{m m''}(n)
\stackrel{\circ}{\varphi}_{m''s}(\lambda,n) \nonumber\\
=\lambda\stackrel{\circ}{\varphi}_{ms}(\lambda,n).
\end{eqnarray}
As a first boundary condition, we can change the zeroth condition to
a more general one corresponding to the homogeneous boundary conditions
$ (\varphi_{ms}(-1)-\varphi_{ms}(0))/\Delta={\cal D}_s\varphi_{ms}(0)$.
In what follows we assume the step of finite-difference differentiation
$\Delta $ to equal 1. According to the conditions (\ref{1.3}), the
functions $\varphi_{ms}(n)$ and $\stackrel{\circ}{\varphi}_{ms}(n)$ are
equal to zero on the lines $(-1, m)$, $(0, m)$ except for the points
with $m=s$ on the line $n=0$ where
$\varphi_{ss}(n)=\stackrel{\circ}{\varphi}_{ss}(n)=1$.
 From these values of $\varphi_{s}$ and $\stackrel{\circ}{\varphi}_{s}$
one can find the functions $\varphi_{ms}(n)$ and
$\stackrel{\circ}{\varphi}_{ms}(n)$ at all other points of the coordinate
network by using the recurrence equation (\ref{1.1}).
Since the matrix equation (\ref{1.2}) is tridiagonal in the variable $n$,
and owing to the boundary conditions (\ref{1.3}), the vectors
$\Phi_s(n)\equiv\{...\varphi_{0s}(n),\varphi_{1s}(n),...\varphi_{ms}(n)...\}$
of solutions $\Phi(n) =(\Phi_0(n), \Phi_1(n),...\Phi_s(n),..) $
in the spectral variable $\lambda$ are polynomials of the $n$th degree.
At $n=1$ the vector $\Phi_s(1)$ is a polynomial of the first degree;
at $n=2$, the vector $\Phi_s(2)$ is a polynomial in $\lambda$ of the
second degree, and so on. It appears, however, that different elements
$\varphi_{ms}(n,\lambda)$ of the vector of solutions $\Phi_s(n)$ have
different degrees of polynomials. Because of the three-point coupling
with respect to $m$ the functions $\varphi_{ms}(n,\lambda)$ are
polynomials in $\lambda $ of the degree $n-|s-m|$. The maximum $n$th
degree belongs to the elements with $m=s$ and the elements
$\phi_{ms}(n,\lambda)$ vanish at all $s$ lying out of the values
$(n-m, n+m)$. The distribution of polynomial degrees is shown in Fig.1.

\begin{minipage}{5cm}
\begin{figure}[htp]
\centering
\mbox{\epsfxsize=3cm\epsffile{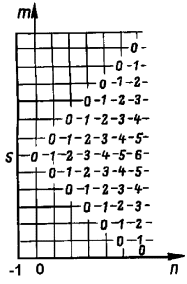}}
\caption{Nonzeroth $\phi_{ms}(n,\lambda)$ are inside and on
boundaries of the cone with its vertex at $(0,s)$;
the numbers at the nodes of the net indicate
the degrees of $\varphi_{ms}(n)$ polynomials.}
\label{fig:1}
\end{figure}
\end{minipage}
\hspace*{.05\textwidth }\hspace*{-0.05em}
\parbox[h,t]{.45\textwidth }{
It is clear that the vectors of solutions,
whose elements are polynomials in $\lambda$ can be orthogonalized.
Berezanskii has shown \cite{berez} that if the elements $a_{nm},
b_{nm}$ and $c_{nm}$ in (\ref{1.1}) are real and $a_{nm}>0$ ($n=1,2,...,
m=...-1,0,1,...)$, one can
introduce polynomials of the first kind ($\varphi_{ms}(\lambda,n)=
P_{s;(n,m)}(\lambda) $) which are orthogonal with the weight of the
spectral matrix  $\rho(\lambda)=\{\rho_{ss'}(\lambda)\}$
}
\\
\begin{eqnarray}
\sum_{ss'=-\infty}^{\infty}\int\varphi_{ms}(\lambda,n)d\rho_{ss'}(\lambda)
\varphi_{s'm'}(\lambda,n')=
\delta_{nn'}\delta_{mm'}
\label{1.6}
\end{eqnarray}
and similarly for $\stackrel{\circ}{\varphi}_{ms}(\lambda,n)$
\begin{eqnarray}
\sum_{ss'}\int\stackrel{\circ}{\varphi}_{ms}(\lambda,n)
d\stackrel{\circ}{\rho}_{ss'}(\lambda)
\stackrel{\circ}{\varphi}_{s'm'}(\lambda,n')=\delta_{nn'}\delta_{mm'}.
\label{1.7}
\end{eqnarray}
The spectral matrix elements $\rho_{ss'}$ (or
$\stackrel{\circ}{\rho}_{ss'}$) are determined by the boundary values
of the special solutions $\psi(n,m)$ (or $\stackrel{\circ}{\psi}(n,m))$
obeying the zeroth condition $\psi(-1,m)=\psi_{m}(-1)=0$ $(m=..-1,0,1...)$.
For example, for $p$ states of the discrete spectrum, the matrix
$\rho(\lambda)$ is the sum of $p$ terms
formed from productions of the column vectors
$\Gamma(\lambda_{\nu})\equiv\{\gamma_s(\lambda_{\nu})\}$ and
the row vectors
$\Gamma^{\dagger}(\lambda_{\nu})\equiv(\gamma_{s}(\lambda_{\nu}))$
$$\rho_{ss'}(\lambda)= \sum_{\nu=1}^{p}\theta(\lambda-\lambda_{\nu})
\gamma_s(\lambda_{\nu})\gamma_{s'}(\lambda_{\nu}), $$
where the elements $\gamma_s(\lambda_{\nu})=\psi_{s}(0,\lambda_{\nu})$
are defined by $\psi(n=0,m)$ and
$\theta(\lambda-\lambda_{\nu})$ is the Heaviside step function equal to 1
at $\lambda=\lambda_{\nu}$ and to zero when  $\lambda\ne\lambda_{\nu}$.
A solution $\psi(n,m)$ to eq.(\ref{1.1}) with the zeroth condition
(or a more general homogeneous condition )\footnote{ Two solutions of the
same equation of second order differ from each other by a normalization
factor at those points $\lambda$ where both of them exist when one of the
boundary conditions is the same.}
can be obtained by multiplying the matrix
$\Phi(\lambda,n)$ by the vector $\Gamma(\lambda)=\Psi(0,\lambda)$:
\begin{eqnarray}
\label{1.5}
 \psi(n,m,\lambda)=\sum_{s=-\infty}^{\infty}
\varphi_{ms}(n,\lambda)\psi_s(0,\lambda).
\end{eqnarray}
At every fixed $n$ and $m$ the summation over $s$ is finite owing to
$\varphi_{ms}(n,\lambda)=0$ outside the interval $(n-m, n+m)$.

\subsection{A discrete version of the Gelfand--Levitan inverse \\ problem}

Using the procedure of orthogonalization of polynomials we construct
unknown polynomial solutions $\varphi_{ms}(\lambda,n)$ normalized
with the spectral weight $\rho_{ss'}(\lambda)$
as a linear combination of the known polynomial solutions
$\stackrel{\circ}{\varphi}_{ms}(\lambda,n)$, orthogonal
with respect to the measure $\stackrel{\circ}{\rho}_{ss'}(\lambda)$
\begin{eqnarray}
\varphi_{ms}(\lambda,n)=
\sum_{n'=0}^{n}\sum_{m'=m-(n-n')}^{m+(n-n')}
 K(n,m;n',m')\stackrel{\circ}{\varphi}_{m's}(\lambda,n').
\label{1.4}
\end{eqnarray}
As one can see above, the validity of this relation is a consequence
of the fact that both the functions,
$\stackrel{\circ}{\varphi}_{ms}(\lambda,n)$ and $\varphi_{ms}(\lambda,n)$,
obey the same discrete Schr\"odinger equation (\ref{1.1})
and the same boundary conditions (\ref{1.3}),
given on the lines $n=-1$ and $n=0$.
Therefore, the solutions $\varphi_{ms}(\lambda,n)$ and
$\stackrel{\circ}{\varphi}_{ms}(\lambda,n)$ are polynomials in $\lambda$
of the same degree $n-|s-m|$, but with different coefficients.
The matrix of the polynomial solutions $\Phi(\lambda,n)$,
which is of degree $n$, is orthogonal to every polynomial matrix of
degree lower than $n$ and hence
to every $\stackrel{\circ}{\Phi}(\lambda,n')$ for $n'<n$
\begin{eqnarray}
\sum_{ss'}\int \varphi_{ms}(\lambda,n) d\rho_{ss'}(\lambda)
\stackrel{\circ}{\varphi}_{s'm'}(\lambda,n')=0.
\label{1.4a}
\end{eqnarray}

Equation (\ref{1.4}) is a discrete analog of the Volterra integral
equation, in which the coefficients $K(n,m;n',m')$ are determined by
the condition of orthogonality of the vector-functions $\Phi(\lambda,n)=(...
\Phi_{1}(\lambda,n),\Phi_{2}(\lambda,n),...,\Phi_{s}(\lambda,n),...)$
orthogonal with the spectral measure $\rho(\lambda)$
to the functions  $\stackrel{\circ}{\Phi}(\lambda,n')$ orthogonal
with the weight matrix $\stackrel{\circ}{\rho}(\lambda)$, when  $n'\leq n$
\begin{eqnarray}
\sum_{ss'}\int \varphi_{ms}(\lambda,n)
(d\stackrel{\circ}{\rho}_{ss'}(\lambda)-d\rho_{ss'}(\lambda))
\stackrel{\circ}{\varphi}_{s'm'}(\lambda,n')= K(n,m;n',m').
\label{1.6a}
\end{eqnarray}
The Volterra equations  (\ref{1.4}) have a triangular form,
$K(n,m;n',m')=0$ for $n'>n$. It is easy to see from (\ref{1.4})
and orthogonality of $\stackrel{\circ}{\varphi}_{sm}(\lambda,n)$
(\ref{1.7}) that   for $ n'<n $
$$K(n,m;n',m')=
\sum_{ss'}\int \varphi_{ms}(\lambda,n)
d\stackrel{\circ}{\rho}_{ss'}(\lambda)
\stackrel{\circ}{\varphi}_{s'm'}(\lambda,n').  $$

Inserting (\ref{1.4}) into (\ref{1.6a}) for $n'< n$ we obtain
the following system of equations
for the orthogonalization coefficients $K(n,m;n',m')$
\begin{eqnarray}
\label{gl}
&&K(n,m;n',m') + K(n,m;n,m)Q(n,m;n',m')+ \\
&&+\sum_{n''=0}^{n-1}\sum_{m''=m-(n-n'')}^{m+(n-n'')}
 K(n,m;n'',m'')Q(n'',m'';n',m')=0, \nonumber
\end{eqnarray}
where
\begin{eqnarray}
\label{Q}
Q(n,m;n',m')=\sum_{ss'}\int
\stackrel{\circ}{\varphi}_{ms}(\lambda,n)(d\rho_{ss'}(\lambda)-
d\stackrel{\circ}{\rho}_{ss'}(\lambda))
\stackrel{\circ}{\varphi}_{s'm'}(\lambda,n').
\end{eqnarray}
Equation (\ref{gl}) is a two-dimensional analog of the Gelfand-Levitan
integral equations for finite-difference equation (\ref{1.1}).
 However, this system is not sufficient to determine the coefficients
 $K$. A supplementary system of equations is obtained upon substituting
 (\ref{1.4}) into the complete relation (\ref{1.6})  at $n'=n$ and $m'=m$
\begin{eqnarray}
&&K^{-2}(n,m;n,m)= Q(n,m;n,m)+
 \nonumber\\
&&+ \sum_{n''=0}^{n-1}\sum_{m''=m-(n-n'')}^{m+(n-n'')}
K^{-1}(n,m;n,m)K(n,m;n'',m'')Q(n'',m'';n,m).
\label{1.7a}
\end{eqnarray}
Now let us find connections between the potential coefficients and
orthogonalization coefficients $K(n,m;n',m')$. To this end, we insert
expression (\ref{1.4})
for the polynomial functions $\varphi_{ms}(\lambda,n)$ in terms of
$\stackrel{\circ}{\varphi}_{ms}(\lambda,n)$ into the Schr\"odinger
difference equation (\ref{1.1})
\begin{eqnarray}
\label{1.8}
&& a_{n m}\sum_{n'=0}^{n-1}\sum_{m'=m-(n-1-n')}^{m+(n-1-n')}
K(n-1,m;n',m')\stackrel{\circ}{\varphi}_{m's}(\lambda,n')\nonumber\\
&& +a_{n+1 m}\sum_{n'=0}^{n+1}\sum_{m'=m-(n+1-n')}^{m+(n+1-n')}
K(n+1,m;n',m')\stackrel{\circ}{\varphi}_{m's}(\lambda,n')\nonumber \\
&& +\sum_{m''=m-1}^{m+1}V_{mm''}(n)\sum_{n'=0}^{n}
\sum_{m'=m''-(n-n')}^{m''+(n-n')}
K(n,m'';n',m')\stackrel{\circ}{\varphi}_{m's}(\lambda,n')\nonumber \\
&&=\lambda
\sum_{n'=0}^{n}\sum_{m'=m-(n-n')}^{m+(n-n')}
K(n,m;n',m')\stackrel{\circ}{\varphi}_{m's}(\lambda,n').
\end{eqnarray}
We transform the r.h.s.
of the above relation by means of the substitution of
$ \lambda\stackrel{\circ}{\varphi}_{m's}(\lambda,n)$ from the finite
difference equation (\ref{1.1a}) for
$\stackrel{\circ}{\varphi}_{ms}(\lambda,n)$
with the known potentials $\stackrel{\circ}{a}_{nm},
\stackrel{\circ}{V}_{mm'}(n)$, $(m'=m-1,m,m+1)$
\begin{eqnarray}
\label{1.8a}
&&\lambda
\sum_{n'=0}^{n}\sum_{m'=m-(n-n')}^{m+(n-n')}
K(n,m;n',m')\stackrel{\circ}{\varphi}_{m's}(\lambda,n')= \nonumber\\
&&+\sum_{n'=0}^{n}\sum_{m'=m-(n-n')}^{m+(n-n')}K(n,m;n',m')
\bigr(\stackrel{\circ}{a}_{n'm'}\stackrel{\circ}{\varphi}_{m's}(\lambda,n'-1)
\nonumber\\
&&+\stackrel{\circ}{a}_{n'+1 m'}\stackrel{\circ}{\varphi}_{m's}(\lambda,n'+1)
+\sum_{m''=m'-1}^{m'+1}
\stackrel{\circ}{V}_{m'm''}(n')\stackrel{\circ}{\varphi}_{m''s}(\lambda,n')
\bigl).
\end{eqnarray}
Further, we take advantage of the orthogonality relation (\ref{1.7}) for
the matrix functions $\stackrel{\circ}{\Phi}(\lambda,n)$
orthogonal with the weight matrix $\stackrel{\circ}{\rho}(\lambda)$.
Multiplying expression (\ref{1.8}) with its transformed r.h.s. (\ref{1.8a})
by $\stackrel{\circ}{\varphi}_{s'm}(\lambda,n+1) $, integrating
over $d\stackrel{\circ}{\rho}_{ss'}(\lambda)$, and summing up over the
indices $s$ and $s'$, we arrive at the relationship between the potentials
$a_{nm}, \stackrel{\circ}{a}_{nm}$ and the coefficients $K(n,m;n'm')$
\begin{eqnarray}
a_{n+1m}= \stackrel{\circ}{a}_{n+1m}\frac{K(n,m;n,m)}{K(n+1,m;n+1,m)}.
\label{1.9}
\end{eqnarray}
The relations for the coefficients $c_{nm}$ and $b_{nm}$ are established
in a similar manner. To determine $b_{nm+1}=V_{mm+1}(n)$, eq.(\ref{1.8})
is multiplied by $\stackrel{\circ}{\varphi}_{s'm+1}(\lambda,n)$ with
(\ref{1.8a}) taken into account and integrated with the weight
$\stackrel{\circ}{\rho}_{ss'}(\lambda)$ by using the orthogonality
(\ref{1.7}). As a result, we have
\begin{eqnarray}
\label{1.11}
b_{nm+1}=\stackrel{\circ}{b}_{nm+1}\frac{K(n,m;n,m)}{K(n,m+1;n,m+1)}
+\stackrel{\circ}{a}_{n m+1}\frac{K(n,m;n-1,m+1)}{K(n,m+1;n,m+1)}\nonumber\\
-a_{n+1 m}\frac{K(n+1,m;n,m+1)}{K(n,m+1;n,m+1)}.
\end{eqnarray}
The relation for $c_{nm}=V_{mm}(n)$ is derived analogously, only
(\ref{1.8}) is multiplied by $\stackrel{\circ}{\varphi}_{s'm}(\lambda,n)$
\begin{eqnarray}
\label{1.12}
c_{nm} =\stackrel{\circ}{c}_{nm}
+\stackrel{\circ}{a}_{n m}\frac{K(n,m;n-1,m)}{K(n,m;n,m)}
-  a_{n+1m}\frac{K(n+1,m;n,m)}{K(n,m;n,m)}.
\end{eqnarray}
Substituting (\ref{1.9}) into (\ref{1.12}) we arrive at
\begin{eqnarray}
\label{1.12a}
c_{nm} = \stackrel{\circ}{c}_{nm}
+ \stackrel{\circ}{a}_{n m}\frac{K(n,m;n-1,m)}{K(n,m;n,m)}
- \stackrel{\circ}{a}_{n+1 m}\frac{K(n+1,m;n,m)}{K(n+1,m;n+1,m)}.
\end{eqnarray}
At $a_{nm}=\stackrel{\circ}{a}_{n m}=1$,
$b_{nm}=\stackrel{\circ}{b}_{n m} =1$, the derived generalized
expressions turn into more simple ones presented in \cite{book}.
The two-dimensional finite-difference
inverse problem under consideration is also a generalization of that
\cite{thesis} with the potential coefficients $a_{nm}\ne 1$, $b_{nm}\ne 1$,
connected nevertheless in a special way.

In principle, it is easy to formulate the problem of restoring the
matrix $V_{mm'}(n)$ in (\ref{1.2}) with all nonzeroth elements,
like in a multichannel problem \cite{book}.
The latter would correspond to the potential being nonlocal with
respect to one of the coordinate variables (in our case $"m"$). If the
consideration were made in the polar coordinate system, nonlocality
with respect to discrete angles would occur. In the case of continuous
coordinates, the inverse problem for the potential, nonlocal relative
to angles, was considered by Kay and Moses \cite{kay}.

\section{Bargmann--Darboux transformations for the \\
two-dimensional discrete Schr\"odinger equation }

In this section, we describe the algebraic procedure
by taking into considaration simple kernels $Q$
in the form of a sum  of several terms with a factorized coordinate
dependence
\begin{eqnarray}
Q(n,m;n',m')=\sum_{\mu=1}^{p}\stackrel{\circ}{\psi}(\lambda_{\mu},n,m)
\stackrel{\circ}{\psi}(\lambda_{\mu},n',m') \nonumber\\
=\sum_{\mu=1}^{p}\sum_{s}\stackrel{\circ}{\varphi}_{ms}(\lambda_{\mu},n)\gamma_s(\lambda_{\mu})
\sum_{s'}\gamma_{s'}(\lambda_{\mu})
\stackrel{\circ}{\varphi}_{s'm'}(\lambda_{\mu},n').
\label{1.Q}
\end{eqnarray}
Here the functions $\stackrel{\circ}{\psi}(\lambda_{\mu},n,m)
\equiv\stackrel{\circ}{\psi}_{m}(\lambda_{\mu},n)$ are combined
as elements of the vector $\stackrel{\circ}{\Psi}(\lambda_{\mu},n)=
(...\stackrel{\circ}{\psi}_{1}(n),\stackrel{\circ}{\psi}_{2}(n),...,
\stackrel{\circ}{\psi}_{m}(\lambda_{\mu},n),...)^{\dagger}$
obtained as a product of the matrix solutions
$\stackrel{\circ}{\Phi}(\lambda_{\mu},n)$ taken at eigenvalues
$\lambda=\lambda_{\mu}$ of the reconstructed $H$ and the vector
$\Gamma(\lambda_{\mu})$
$$\stackrel{\circ}{\psi}(\lambda_{\mu},n,m)=\sum_{s}
\stackrel{\circ}{\varphi}_{ms}(\lambda_{\mu},n)\gamma_s(\lambda_{\mu}). $$
The elements $\gamma_s$ form the normalization matrix
$C(\lambda_{\mu})=
\Gamma(\lambda_{\mu})\Gamma^{\dagger}(\lambda_{\mu})$ with elements
$C_{ss'}(\lambda_{\mu})=
\gamma_s(\lambda_{\mu})\gamma_{s'}(\lambda_{\mu})$
corresponding to the bound state
$\psi_{\mu}(n,m)\equiv\psi(\lambda_{\mu},n,m)$, $\mu=1,2,..p$.

Like $Q$, the ortogonalization kernel $K(n,m;n'm')$ is also presented as
a sum of several factorized terms.
Really, substituting  (\ref{1.Q}) into the Gelfand Levitan equation
(\ref{gl}), we obtain
\begin{eqnarray}
\label{Knm}
&&K(n,m;n',m') =\\
&&-\sum_{\mu=1}^{p}\{\sum_{n''=0}^{n}\sum_{m''=m-(n-n'')}^{m+(n-n'')}
 K(n,m;n'',m'') \stackrel{\circ}{\psi}(\lambda_{\mu},n'',m'')\}
\stackrel{\circ}{\psi}(\lambda_{\mu},n',m'). \nonumber
\end{eqnarray}
Noting that the expression in braces is the solution $\psi_{\mu}(n,m)$
(\ref{1.4}) at $\lambda=\lambda_{\mu} $ of eq.(\ref{1.1}) with the desired
potentials $a_{nm}, b_{nm}$ and $c_{nm}$, we immediately receive
\begin{eqnarray}
K(n,m;n',m')=-\sum_{\mu=1}^{p}\psi_{\mu}(n,m)
\stackrel{\circ}{\psi}(\lambda_{\mu},n',m').
\label{1.K}
\end{eqnarray}
It is evident now that the new wave functions $\varphi_{ms}(\lambda,n)$,
determined by (\ref{1.4}) with the kernel $K$ taken in the form (\ref{1.K}),
are related to the old ones $\stackrel{\circ}{\varphi}_{ms}(\lambda,n)$ by
\begin{eqnarray}
\varphi_{ms}(\lambda,n)=-
\sum_{\mu}^{p}\psi_{\mu}(n,m)\sum_{n'=0}^{n}\sum_{m'=m-(n-n')}^{m+(n-n')}
 \stackrel{\circ}{\psi}(\lambda_{\mu},n',m')
\stackrel{\circ}{\varphi}_{m's}(\lambda,n').
\label{1.4bar}
\end{eqnarray}

In view of (\ref{1.K}) for $K(n,m;n',m')$ in eqs. (\ref{1.9}),
(\ref{1.11}) and (\ref{1.12}), one can immediately write expressions
for discrete potentials in the closed form
\begin{eqnarray}
\label{a-nm}
a_{n+1m}= \stackrel{\circ}{a}_{n+1m}
\frac{\sum_{\mu=1}^{p}\psi_{\mu}(n,m)
\stackrel{\circ}{\psi}(\lambda_{\mu},n,m)}
{\sum_{\mu=1}^{p}\psi_{\mu}(n+1,m)
\stackrel{\circ}{\psi}(\lambda_{\mu},n+1,m)};
\end{eqnarray}
\begin{eqnarray}
\label{b-nm}
b_{nm+1}&=&\stackrel{\circ}{b}_{nm+1}\frac{
\sum_{\mu=1}^{p}\psi_{\mu}(n,m)
\stackrel{\circ}{\psi}(\lambda_{\mu},n,m)}
{\sum_{\mu=1}^{p}\psi_{\mu}(n,m+1)
\stackrel{\circ}{\psi}(\lambda_{\mu},n,m+1)}\nonumber\\
&+&\stackrel{\circ}{a}_{n,m+1}
\frac{
\sum_{\mu=1}^{p}\psi_{\mu}(n,m)
\stackrel{\circ}{\psi}(\lambda_{\mu},n-1,m+1)}
{\sum_{\mu=1}^{p}\psi_{\mu}(n,m+1)
\stackrel{\circ}{\psi}(\lambda_{\mu},n,m+1)}
\nonumber\\
&-& a_{n+1 m}\frac{\sum_{\mu=1}^{p}\psi_{\mu}(n+1,m)
\stackrel{\circ}{\psi}(\lambda_{\mu},n,m+1)}
{ \sum_{\mu=1}^{p}\psi_{\mu}(n,m+1)
\stackrel{\circ}{\psi}(\lambda_{\mu},n,m+1)}
\end{eqnarray}
and
\begin{eqnarray}
\label{c-nm}
&& c_{nm} = \stackrel{\circ}{c}_{nm}
+\stackrel{\circ}{a}_{n m}
\frac{\sum_{\mu=1}^{p}\psi_{\mu}(n,m)
\stackrel{\circ}{\psi}(\lambda_{\mu},n-1,m)}
{ \sum_{\mu=1}^{p}\psi_{\mu}(n,m)
\stackrel{\circ}{\psi}(\lambda_{\mu},n,m)}\\
&&- \frac{ \sum_{\mu=1}^{p}\stackrel{\circ}{a}_{n+1 m}\psi_{\mu}(n+1,m)
\stackrel{\circ}{\psi}(\lambda_{\mu},n,m)}
{\sum_{\mu=1}^{p}\psi_{\mu}(n+1,m)
\stackrel{\circ}{\psi}(\lambda_{\mu},n+1,m)}. \nonumber
\end{eqnarray}
The solutions $\psi_{\mu}(n,m) $ have to be found
from the Gelfand-Levitan equations (\ref{gl}) and (\ref{1.7a})
taking account of (\ref{1.Q}) and (\ref{1.K}).
Thus, the operator $K(n,m;n',m')$ defined by (\ref{1.K})
transforms the solutions $\stackrel{\circ}{\varphi}_{sm}(n)$ of eq.
(\ref{1.1a}) into the solutions $\varphi_{sm}(n)$ of eq.(\ref{1.1}),
determined by (\ref{1.4}) or (\ref{1.4bar}), with
the potentials $a_{nm}, b_{nm}$ and $c_{nm}$ defined by (\ref{a-nm}),
(\ref{b-nm}) and (\ref{c-nm}).

It is not difficult to see from the definitions (\ref{Q}) and (\ref{1.6a})
that the kernels $Q$ and $K$ like (\ref{1.Q}) and (\ref{1.K})
can be obtained provided that the spectral weight
functions $\rho(\lambda)$ and $\stackrel{\circ}{\rho}(\lambda) $
for both the sets of potentials coincide
except, for instance, $p$ eigenvalues at $\lambda=\lambda_{\mu}$.
This permits one to construct potentials with $p$ new bound states
by using (\ref{a-nm}), (\ref{c-nm}) and (\ref{b-nm}) or generate
the family of spectral-equivalent potentials whose spectra coincide
$\lambda_{\mu}=\stackrel{\circ}{\lambda}_{\mu}$ and it is only
the normalization factors $C_{\mu}\ne \stackrel{\circ}{C}_{\mu}$
that are different. In the latter case $Q$ is taken in the form
\begin{eqnarray}
Q(n,m;n',m')
=\sum_{\mu=1}^{p}\sum_{ss'}\stackrel{\circ}{\varphi}_{ms}(\lambda_{\mu},n)
(C_{ss'}(\lambda_{\mu}) - \stackrel{\circ}{C}_{ss'}(\lambda_{\mu}))
\stackrel{\circ}{\varphi}_{s'm'}(\lambda_{\mu},n')
\label{Q-iso}
\end{eqnarray}
and the above procedure can be used to construct
spectral-equivalent operators $ \stackrel{\circ}{H}$ and $H$.
In spite of a complicated form of the expressions for the potentials
(\ref{a-nm}), (\ref{b-nm}) and (\ref{c-nm}), they are simplified for
a large set of particular cases. For example, if we deal with the free
discrete Schr\"odinger equation as a reference one, fixed by the choice
$\stackrel{\circ}{c}_{nm}\equiv 0$,
$\stackrel{\circ}{a}_{nm}=\stackrel{\circ}{b}_{nm}\equiv 1$.

Let us now consider another simple case when the Hamiltonians differ
only by spectral data at one bound state.
In this case, the summation over $\mu$ in all formulae
(\ref{1.4bar}) -- (\ref{c-nm}) vanishes. General solutions
$\varphi_{sm}(\lambda,n)$ from (\ref{1.4bar}) at arbitrary
$\lambda$ can be written  as
\begin{eqnarray}
\varphi_{ms}(\lambda,n)=-
\psi(n,m)\sum_{n'=0}^{n}\sum_{m'=m-(n-n')}^{m+(n-n')}
 \stackrel{\circ}{\psi}(n',m')
\stackrel{\circ}{\varphi}_{m's}(\lambda,n'),
\label{1.4b}
\end{eqnarray}
where $\stackrel{\circ}{\psi}(n,m) $ is a special solution of the
Schr\"odinger equation (\ref{1.1a}) for the value of the spectral
parameter $\lambda =\mu $ and $\psi(n,m) $ is the solution of
eq.(\ref{1.1}) with the same eigenvalue $\lambda =\mu $.
New potentials are expressed in terms of the known old
potentials $\stackrel{\circ}{a}_{nm}$, $\stackrel{\circ}{c}_{nm}$,
and $\stackrel{\circ}{b}_{nm}$, functions $\stackrel{\circ}{\psi}(n,m)$
and functions $\psi(n,m)$ that can be determined from  the second
Gelfand-Levitan equation (\ref{1.7a})
\begin{eqnarray}
\label{a-nm1}
a_{n+1m}= \stackrel{\circ}{a}_{n+1m}
\frac{\psi(n,m)\stackrel{\circ}{\psi}(n,m)}
{\psi(n+1,m)\stackrel{\circ}{\psi}(n+1,m)},
\end{eqnarray}
\begin{eqnarray}
\label{b-nm1}
 b_{nm+1} &=&
\frac{\psi(n,m)}
{\psi(n,m+1)}
\Biggl(\stackrel{\circ}{b}_{nm+1}
\frac{
\stackrel{\circ}{\psi}(n,m)}
{\stackrel{\circ}{\psi}(n,m+1)}\nonumber\\
&+&\stackrel{\circ}{a}_{n m+1}
\frac{
\stackrel{\circ}{\psi}(n-1,m+1)}
{\stackrel{\circ}{\psi}(n,m+1)}
- \stackrel{\circ}{a}_{n+1 m} \frac{
\stackrel{\circ}{\psi}(n,m)}
{\stackrel{\circ}{\psi}(n+1,m)}\Biggr).
\end{eqnarray}
and
\begin{eqnarray}
\label{c-nm1}
&& c_{nm} = \stackrel{\circ}{c}_{nm}
+\stackrel{\circ}{a}_{n m}
\frac{\stackrel{\circ}{\psi}(n-1,m)}
{\stackrel{\circ}{\psi}(n,m)}
- \stackrel{\circ}{a}_{n+1 m} \frac{
\stackrel{\circ}{\psi}(n,m)}
{\stackrel{\circ}{\psi}(n+1,m)}.
\end{eqnarray}
It should be noted that relationships between potentials and functions
can be obtained within the Darboux transformation method or factorised
method without using formulae of the inverse problem.

{\it Connection between Darboux transformations and inverse problem ones.}
It is interesting to note that transformation (\ref{1.4b})
with one bound state corresponds to Darboux transformation for
the finite-difference equation (\ref{1.1}).
Indeed, let us search for a solution $\varphi_{ms}(\lambda,n)$
of eq.(\ref{1.1}) with some initially unknown potentials in the form
(\ref{1.4b}). Next it is necessary to find conditions for the potentials
$a_{nm}, b_{nm}$ and $c_{nm}$ and special functions $\psi(n,m)$ at which
general solutions $ \varphi_{ms}(\lambda,n) $ specified by (\ref{1.4b})
will satisfy the discrete Schr\"odinger equation (\ref{1.1}).
Substitute (\ref{1.4b}) into (\ref{1.1})
\begin{eqnarray}
\label{cond}
&& a_{n+1 m}\psi(n+1,m)\sum_{n'=0}^{n+1}\sum_{m'=m-(n+1-n')}^{n+1-n'}
\stackrel{\circ}{\psi}(n',m')
\stackrel{\circ}{\varphi}_{m's}(\lambda,n') \nonumber\\
&& + a_{n m}\psi(n-1,m)\sum_{n'=0}^{n-1}\sum_{m'=m-(n-1-n')}^{m+(n-1-n')}
\stackrel{\circ}{\psi}(n',m')
\stackrel{\circ}{\varphi}_{m's}(\lambda,n')\nonumber\\
&& +\sum_{m''=m-1}^{m+1}V_{mm''}(n)\psi(n,m'')
\sum_{n'=0}^{n}\sum_{m'=m''-(n-n')}^{m''+(n-n')}\stackrel{\circ}{\psi}(n',m')
\stackrel{\circ}{\varphi}_{m's}(\lambda,n')= \nonumber\\
&&=\lambda\psi(n,m)\sum_{n'=0}^{n}\sum_{m'=m-(n-n')}^{m+(n-n')}
\stackrel{\circ}{\psi}(n',m')
\stackrel{\circ}{\varphi}_{m's}(\lambda,n').
\end{eqnarray}
Transform the r.h.s. of (\ref{cond}) substituting
$\lambda\stackrel{\circ}{\varphi}_{m's}(\lambda,n)$ from (\ref{1.1a})
\begin{eqnarray}
\label{rhs}
&&\lambda\psi(n,m)\sum_{n'=0}^{n}\sum_{m'=m-(n-n')}^{m+(n-n')}
\stackrel{\circ}{\psi}(n',m')
\stackrel{\circ}{\varphi}_{m's}(\lambda,n')= \nonumber\\
&&\lambda\psi(n,m)\sum_{n'=0}^{n}\sum_{m'=m-(n-n')}^{m+(n-n')}
\stackrel{\circ}{\psi}(n',m')[
\stackrel{\circ}{a}_{n'm'}\stackrel{\circ}{\varphi}_{m's}(\lambda,n'-1)+
\stackrel{\circ}{a}_{n'+1 m'}\stackrel{\circ}{\varphi}_{m's}(\lambda,n'+1)
\nonumber\\
&&+\stackrel{\circ}{c}_{n'm'}
\stackrel{\circ}{\varphi}_{m's}(\lambda,n')+
\stackrel{\circ}{b}_{n'm'}\stackrel{\circ}{\varphi}_{m'-1 s}(\lambda,n')+
\stackrel{\circ}{b}_{n m'+1}\stackrel{\circ}{\varphi}_{m'+1 s}(\lambda,n')].
\end{eqnarray}
Further, to obtain the relations for potentials $a_{n+1m}$, $b_{nm+1}$
and $c_{nm}$, multiply eq.(\ref{cond}) with its transformed r.h.s.
(\ref{rhs}) by $\stackrel{\circ}{\varphi}_{m s}(\lambda,n+1)$,
$\stackrel{\circ}{\varphi}_{m+1 s}(\lambda,n)$ and
$\stackrel{\circ}{\varphi}_{m s}(\lambda,n)$ and take into considaration
the completeness relation (\ref{1.7}) for the functions
$\stackrel{\circ}{\varphi}(\lambda)$.
The expressions for $a_{nm}$, $b_{nm}$ and $c_{nm}$ thus
derived coincide with formulae (\ref{a-nm1}), (\ref{b-nm1})
and (\ref{c-nm1}), correspondingly, obtained from the formulae of the
inverse problem.

\section{Conclusion}

The Gelfand-Levitan spectral inverse problem for the discrete two-dimensional
\break
Schr\"odinger equation is considered on the basis of the Berezanskii
technique of orthogonalization of polynomial matrices.
By using the derived formulae of the inverse problem, discrete
Bargmann--Darboux transformations in two dimensions are given.
 Analytic relationships are established between the solutions for
two different sets of discrete potentials and the potentials themselves.
\\

I thank Dr.V.M.Muzafarov and E.P.Velicheva for helpful discussions.\\
A part of this work was done in June of 1998 at the Erwin Schr\"odinger
International Institute for Mathematical Physics, Wien, Austria.

{}

\end{document}